\documentclass{amsart}
\UseRawInputEncoding
\usepackage{graphics, graphicx}
\usepackage[english]{babel}
\usepackage{amsfonts}
\usepackage{amsfonts,amssymb,amsmath,amsbsy,amsthm,amscd}
 \usepackage{hyperref}
\newtheorem{theorem}{Theorem}

\def\dfrac#1#2{\displaystyle{#1\over #2}}
\def\za{\alpha}
\def\zb{\beta}

\def\zt{\theta}
\def\bv{{\bf v}}
\def\bp{{\bf p}}

\def\zr{\rho}

\def\Div{\mbox{div}\,}

\def\bB{{\bf B}}

\def\bE{{\bf E}}

\begin{document}

\markboth{Rozanova, Chizhonkov}{Relativistic cold plasma on any given time interval}

%
%

\title[Relativistic cold plasma on any given time interval]{
A sufficient condition for the existence of smooth solutions of the relativistic cold plasma equations on any given time interval
}

\author{Olga S. Rozanova*}

\address{ Mathematics and Mechanics Department, Lomonosov Moscow State University, Leninskie Gory,
Moscow, 119991,
Russian Federation,
rozanova@mech.math.msu.su}

\author{Evgeniy V. Chizhonkov}

\address{ Mathematics and Mechanics Department, Lomonosov Moscow State University, Leninskie Gory,
Moscow, 119991,
Russian Federation,
chizhonk@mech.math.msu.su}

\subjclass{Primary 35Q60; Secondary 35L60, 35L67, 34M10}

\keywords{Quasilinear hyperbolic system,
relativistic plasma oscillations,  smooth solutions, blow up}

\maketitle


\begin{abstract}
In terms of initial data, a sufficient condition for the smoothness of the solution to the Cauchy problem for one-dimensional relativistic cold plasma equations over any given time interval is found. Unlike the nonrelativistic case, such sufficient conditions take into account the smallness properties of not only the derivatives of the initial data but also the initial data themselves. The accuracy of the obtained initial condition is investigated using a numerical experiment. The structure of the emerging singularities is also studied.
 \end{abstract}


\section{Introduction}	
The excitation of oscillations and waves in plasma using short, powerful laser pulses has sparked interest in new mathematical problem formulations. This primarily concerns the study of relativistic effects associated with the dynamics of light particles (electrons)~\cite{ESL09}, \cite{BEH16}. Plasma is a highly nonlinear medium in which even relatively small initial collective particle displacements lead to the excitation of oscillations and waves of sufficiently large amplitude. Therefore, the study of the properties of nonlinear oscillations and waves in plasma, as an important area of nonlinear physics, is not only of academic interest but also relates to the potential for their practical use in accelerating charged particle bunches to high energies and generating terahertz radiation pulses~\cite{ESL09}. In a plasma without dissipation, the evolution of highly nonlinear oscillations and waves leads to their breakdown
due to the emergence of an electron density singularity~\cite{david72}. As for plasma oscillations, the question of their time evolution and reversal was first considered in the work of \cite{Daw59}, where the time of intersection of electron trajectories in flat, cylindrical and spherical geometry was found; the corresponding numerical experiments can be found  in \cite{CH18}.
  The intersection of particle trajectories in a Lagrangian description of the medium is equivalent to the emergence of a density singularity in the Euler variables \cite{ZM}.
  The influence of relativistic effects on the evolution of Langmuir oscillations in planar geometry is considered in~\cite{FCh2014}, where the breaking time is found and it is shown that a density singularity can arise for arbitrarily small initial deviations of the density from the equilibrium value due to the dependence of the oscillation frequency on their amplitude.

  However, from a practical point of view, when the existence of regular oscillations must be of sufficiently long duration, it is important to know what initial data should be chosen for this, and even more important to associate the initial data with a predetermined time interval over which it is desired to ensure a solution with regular behavior.

In previous publications, the question of the blow-up time was studied for small deviations from the equilibrium using asymptotic methods. For this reason, initial perturbations of a special form, primarily with practical applications, were investigated. In the present paper, arbitrary initial functions are considered. Another important aspect of this paper is consideration of the blow-up time of relativistic oscillations from two points of view.
On the one hand, the existence time of a smooth solution can be estimated based on information about the initial data. Conversely, the initial data can be selected such that a smooth solution exists for a predetermined time. In comparison, the methods of ~\cite{RCh2021} in the general case could not answer the question about the period of time of existence of a smooth solution.

It should be noted that, from a mathematical point of view, the fundamental absence of nontrivial smooth solutions to the cold plasma equations is a consequence of the difference in the periods of oscillations along various Lagrangian particle trajectories \cite{RozPhD2024}. In this sense, the pattern of the emergence of singularities in the solution of equations describing relativistic plasma is similar to the pattern of the emergence of singularities in the solution of equations with cylindrical and axial symmetry of solutions of non-relativistic three-dimensional plasma. (If we consider axisymmetric solutions with radial symmetry of the abstract system of equations of non-relativistic cold plasma in a space of arbitrary dimension, then the situation differs only in dimensions 1 and 4 \cite{Roz_exept}). The idea of ​​estimating the lifetime of a smooth solution for solutions with cylindrical and axial symmetry was implemented in \cite{Roz_MMAS23} (see also \cite{RCh2021_coll} for oscillations with collisions). It is similar to the one used in the present paper, but  the linearization of the characteristic system allows us to formulate sufficient conditions on the initial data, ensuring the smoothness of the solution over a given interval, in a more compact and easily verifiable form. This paper can be considered a continuation of \cite{RCh2021}, and we make substantial use of the results obtained there.

The paper is organized as follows. In Section \ref{S2}, the equations for a one-dimensional relativistic cold plasma are derived.
In Section  \ref{S3} we prove the main result of this paper,  a sufficient condition for the Cauchy initial data to ensure the existence of a smooth solution on any given time interval (Theorem \ref{T1}). The resulting initial condition is analyzed for small initial perturbations.
In Section  \ref{S4}, the accuracy of the sufficient conditions of Theorem \ref{T1} is investigated using a numerical experiment and the nature of the emerging singularity of the solution is studied. Section \ref{S5} discusses the applicability of the sufficient conditions.

\section{Statement of the problem and the main result}\label{S2}
We consider the plasma as a relativistic electron fluid, neglecting recombination effects and ion motion. The
system of {\it a "cold" plasma}, including the hydrodynamic equations together with Maxwell's equations
in vector form is the following (see, for example,~\cite{ABR78},\cite{GR75}):
\begin{equation}
\label{base1}
\begin{array}{l}
\dfrac{\partial n }{\partial t} + \Div(n \bv)=0\,,\quad
\dfrac{\partial \bp }{\partial t} + \left( \bv \cdot \nabla \right) \bp
= e \, \left( \bE + \dfrac{1}{c} \left[\bv \times  \bB\right]\right),\vspace{0.5em}\\
\gamma = \sqrt{ 1+ \dfrac{|\bp|^2}{m^2c^2} }\,,\quad
\bv = \dfrac{\bp}{m \gamma}\,,\vspace{0.5em}\\
\dfrac1{c} \frac{\partial \bE }{\partial t} = - \dfrac{4 \pi}{c} e n \bv
 + {\rm rot}\, \bB\,,\quad
\dfrac1{c} \frac{\partial \bB }{\partial t}  =
 - {\rm rot}\, \bE\,, \quad \Div \bB=0\,,
\end{array}
\end{equation}
where
$e, m$ are the charge and mass of the electron (the electron charge has a negative sign: $e < 0$),
$c$ is the speed of light;
$n, \bp, \bv$ are the concentration, momentum, and velocity
of electrons;
$\gamma$ is the Lorentz factor;
$\bE, \bB$ are vectors the electric and magnetic fields, respectively.

Let us assume that the solution depends on only one spatial coordinate $x$. Then  system~\eqref{base1} can be significantly simplified:
\begin{equation}
\begin{array}{c}
\dfrac{\partial n }{\partial t} +
\dfrac{\partial }{\partial x}
\left(n\, v_x \right)
=0,\quad
\dfrac{\partial p_x }{\partial t} + v_x \dfrac{\partial p_x}{\partial x}= e \,E_x,
\vspace{1 ex}\\
\gamma = \sqrt{ 1+ \dfrac{p_x^2}{m^2c^2}}\,, \quad
{v_x} = \dfrac{p_x}{m \,\gamma}, \quad
\dfrac{\partial E_x }{\partial t} = - 4 \,\pi \,e \,n\, v_x\,.
\end{array}
\label{3gl2}
\end{equation}

If we introduce dimensionless quantities
$$
\rho = k_p x, \quad \theta = \omega_p t, \quad
V = \dfrac{v_x}{c}, \quad
P = \dfrac{p_x}{m\,c}, \quad
E = -\,\dfrac{e\,E_x}{m\,c\,\omega_p}, \quad
N = \dfrac{n}{n_0},
$$
where $\omega_p = \left(4 \pi e^2n_0/m\right)^{1/2}$ is the plasma frequency, $n_0$ is the value of the unperturbed electron density, $k_p = \omega_p /c$,
then  system \eqref{3gl2} takes the form
\begin{equation}
\begin{array}{c}
\dfrac{\partial N }{\partial \theta} +
\dfrac{\partial }{\partial \rho}
\left(N\, V \right)
=0,\quad
\dfrac{\partial P }{\partial \theta} + E +
V \dfrac{\partial P}{\partial \rho} = 0, \vspace{1.5 ex}\\
\gamma = \sqrt{ 1+ P^2}, \quad
V = \dfrac{P}{\gamma},\quad
\dfrac{\partial E }{\partial \theta} = N\, V\,.
\end{array}
\label{3gl3}
\end{equation}
From the first and last equations~\eqref{3gl3} we get
$$
\dfrac{\partial }{\partial \theta}
\left[ N +
\dfrac{\partial }{\partial \rho} E \right] = 0,
$$
therefore, assuming that $N$ and $E$ tend to constants at infinity, we obtain
\begin{equation*}
 N = 1 -
\dfrac{\partial  E }{\partial \rho}.
\label{3gl4}
\end{equation*}
This allows us to eliminate the density component from the system and obtain
equations describing
flat one-dimensional relativistic plasma oscillations~\cite{CH18}
\begin{equation}\label{u1}
\dfrac{\partial P }{\partial \theta}+
V\,\dfrac{\partial P}{\partial \rho}  + E = 0,\quad
\dfrac{\partial E }{\partial \theta}  +
V\,\dfrac{\partial E}{\partial \rho}- V = 0,
\quad V = \dfrac{P}{\sqrt{1+P^2}}\,.
\end{equation}
Here $\rho$ and $\theta$ are dimensionless coordinates in
space and time, respectively. The variable $P$ describes the electron momentum, $V$ is the electron velocity, and $E$ is the electric field strength.

Let us pose for \eqref{u1} the Cauchy problem
with the initial conditions
\begin{equation}\label{cd1}
     P(\rho,0) = P_0(\rho), \quad
     E(\rho,0) = E_0(\rho), \quad
 \rho \in {\mathbb R}.
\end{equation}

The main result of this paper is the following theorem.

Let us denote $C(\rho)=2\sqrt{1+P_0^2}+E_0^2$ and $ K_-=\frac{8}{C^3}$.

\begin{theorem}\label{T1}
Let the initial data \eqref{cd1} belong to the class $C^1(\mathbb R)$ and for these data $C(\rho)$ is not identically constant.
The condition
\begin{equation}\label{condn}
\inf\limits_{\rho\in \mathbb R}\left[K_-^{n-1}(1-e_0)^2-\frac{ e_0^2}{K_-}-p_0^2 \right]>0,
\end{equation}
where  $p_0=P_0'$, $e_0=E_0'$,
ensures that the solution to problem \eqref{u1}, \eqref{cd1}, remains smooth at least
until time $t_*> n \pi$, $n\in \mathbb N$.

In other words, if we need to ensure the smoothness of the solution of the problem \eqref{u1}, \eqref{cd1} up to any time $T>0$, we need to choose initial data that satisfy the condition \eqref{condn} for $n=[{\frac{T}{\pi}}]$.
\end{theorem}

\section{The proof of Theorem \ref{T1} and further theoretical analysis} \label{S3}

\subsection{The proof of Theorem \ref{T1}}
This paper is a direct continuation of \cite{RCh2021} and uses the arguments developed there. We will list them briefly.

The system of equations \eqref{u1} is hyperbolic, therefore it has a smooth solution locally in time, and the formation of a singularity is associated either with the unboundedness of the solution itself or its derivatives \cite{Daf16}. We write the system of characteristics for \eqref{u1}:
\begin{equation}\label{char1}
     \dfrac{dP}{d\theta}=-E,\quad \dfrac{dE}{d\theta}=\dfrac{P}{\sqrt{1+P^2}},\quad \dfrac{d\rho}{d\theta}=\dfrac{P}{\sqrt{1+P^2}}.
     \end{equation}
It follows from this that
\begin{equation*}\label{first_int}
2\sqrt{1+P^2}+E^2=
2\sqrt{1+P^2(\rho_0)}+E^2(\rho_0)=C(\rho_0)\ge 2
\end{equation*}
along
the characteristic starting from the point $\rho_0$. Therefore, the solution itself
remains bounded:
$$
  1\le\sqrt{1+P^2}\le \frac{C(\rho_0)}{2}, \quad 2 + E^2  \le C(\rho_0).
$$

The period $T(\rho_0)$ can be calculated as
$$
  T = 2 \int\limits_{P_-}^{P_+}  \dfrac{dP}{ \sqrt{C-2\sqrt{1+P^2}}},\quad
  P_\pm=\pm\frac{\sqrt{C^2-4}}{2},
$$
For brevity, we omit the $\rho_0$ argument here and below. The period
tends to $2\pi$ as $C\to 2$, but increases as $C$ increases.
Unlike the non-relativistic case, where the period is the same along each
characteristic and equals $2\pi$, in the relativistic
case, the period is unique along each characteristic.

This is the reason why, for any nontrivial initial data of general form, the characteristics will necessarily intersect, and within a finite time, the solution will lose smoothness (see \cite{RozPhD2024}, Lemma 3.2).
However, it is possible to find a time during which the solution is guaranteed to remain smooth.

Let us construct the extended system. To do this, we  denote $p=P_\rho$, $e=E_\rho$ and differentiate \eqref{u1} with respect to $\rho$.
\begin{equation}\label{char1d}
     \dfrac{dp}{d\theta}=-e-\frac{p^2}{(1+P^2)^{3/2}},\quad \dfrac{de}{d\theta}=(1-e)\frac{p}{(1+P^2)^{3/2}}.
     \end{equation}
We see that  system \eqref{char1d} cannot be considered separately from \eqref{char1}. However, if we assume that the equation following from \eqref{char1}
\begin{equation}\label{char1d+}
\dfrac{dP}{d\theta}=- \sqrt{C-2\sqrt{1+P^2}},
\end{equation}
describing the behavior of the bounded function $P$, is solved, then \eqref{char1d} can be considered as a quadratically nonlinear system of two equations with known time-dependent coefficients.

We  only consider the case where the expression
$C(\rho_0)=2\sqrt{1+P_0^2}+E_0^2$ is not identically equal to a constant. The case where it is equal to a constant is special; it corresponds to a simple wave solution and is fully analyzed in \cite{RCh2021}.

For convenience, we introduce the periodic function $K(\theta)=(1+P^2(\theta))^{-3/2}$, defined using \eqref{char1d+}, which can be estimated as
\begin{equation*}\label{estK}
0<K_-=\frac{8}{C^3(\rho_0)} \le K(\theta)\le 1.
\end{equation*}

The system \eqref{char1d} can be linearized using the  Radon's lemma \cite{Radon}. This increases the dimensionality of the system. We present the formulation of this lemma.

\medskip

\begin{theorem}[The Radon lemma]
\label{T2} A matrix Riccati equation
\begin{equation}
\label{Ric}
 \dot W =M_{21}(t) +M_{22}(t)  W - W M_{11}(t) - W M_{12}(t) W,
\end{equation}
 {\rm (}$W=W(t)$ is a matrix $(n\times m)$, $M_{21}$ is a matrix $(n\times m)$, $M_{22}$ is a matrix  $(m\times m)$, $M_{11}$ is a matrix  $(n\times n)$, $M_{12} $ is a matrix $(m\times n)${\rm )} is equivalent to the homogeneous linear matrix equation
\begin{equation}
\label{Lin}
 \dot Y =M(t) Y, \quad M=\left(\begin{array}{cc}M_{11}
 & M_{12}\\ M_{21}
 & M_{22}
  \end{array}\right),
\end{equation}
 {\rm (}$Y=Y(t)$  is a matrix $(n\times (n+m))$, $M$ is a matrix $((n+m)\times (n+m))$ {\rm )} in the following sense.

Let on some interval ${\mathcal J} \in \mathbb R$ the
matrix-function $\,Y(t)=\left(\begin{array}{c}Q(t)\\ P(t)
  \end{array}\right)$ {\rm (}$Q$  is a matrix $(n\times n)$, $P$  is a matrix $(n\times m)${\rm ) } be a solution of \eqref{Lin}
  with the initial data
  \begin{equation*}\label{LinID}
  Y(0)=\left(\begin{array}{c}I\\ W_0
  \end{array}\right)
  \end{equation*}
   {\rm (}$ I $ is the identity matrix $(n\times n)$, $W_0$ is a constant matrix $(n\times m)${\rm ) } and  $\det Q\ne 0$ on ${\mathcal J}$.
  Then
{\bf $ W(t)=P(t) Q^{-1}(t)$} is the solution of \eqref{Ric} with
$W(0)=W_0$ on ${\mathcal J}$.
\end{theorem}

System \eqref{char1d} can be written as \eqref{Ric} with
\begin{eqnarray*}\label{M}
W=\begin{pmatrix}
  u\\
  v
\end{pmatrix},\quad
M_{11}=\begin{pmatrix}
  0\\
\end{pmatrix},\quad
 M_{12}=\begin{pmatrix}
  1 & 0\\
\end{pmatrix},\\
M_{21}=\begin{pmatrix}
 1 & 0 \\
\end{pmatrix},\quad
M_{22}=\begin{pmatrix}
  - 2\,F& -1\\
  1-d\,G  & -d\,F\\
\end{pmatrix}.\\\nonumber
\end{eqnarray*}

\medskip

The result of the linearization is
\begin{equation}\label{char1dlin}
\dfrac{d \bar p}{d \theta}=-\bar e,\quad \dfrac{d \bar e}{d \theta}=\bar p K, \quad \dfrac{d q}{d \theta}=\bar p K,
\end{equation}
with the initial conditions
\begin{equation*}\label{char1d_CD}
\bar p(0)=p_0(\rho_0)=P_0'(\rho_0), \quad \bar e(0)=e_0(\rho_0)=E_0'(\rho_0), \quad q(0)=q_0(\rho_0)=1.
     \end{equation*}
Moreover, along the characteristic starting from $\rho_0$, we have
\begin{equation*}\label{char1dsol}
p(\theta)= \frac{\bar p (\theta)}{q(\theta)}, \quad e(\theta)= \frac{\bar e (\theta)}{q(\theta)}.
\end{equation*}
Thus, the derivatives of the solution on a given characteristic vanish if and only if $q$ vanishes.

The study of linear systems is generally simpler than that of nonlinear systems, and It is numerically easier to determine whether a function vanishes than whether it tends to infinity.

Note that the last two equations \eqref{char1dlin} immediately imply
\begin{equation*}\label{q_e}
\bar e=q+e_0-1,
\end{equation*}
which allows us to eliminate the variable $\bar e$ and reduce the dimension of the system \eqref{char1dlin} to two. Since any instant  can be taken as the initial point in time, instead of $\theta=0$, we further choose the points we need as the zero value.

The solution of \eqref{char1dlin} cannot be found explicitly, but it can be estimated;
namely, we restrict the position of the phase trajectory on the phase plane $(q, \bar p)$.

To do this, we note that since $ K_-\le K(\theta)\le K_+=1$, then, according to Chaplygin's theorem on differential inequalities \cite{Chap}, the integral curve of the equation
\begin{equation}\label{q_p}
\dfrac{d \bar p}{d q}=\frac {-q+1-e_0}{K \bar p}, \quad \bar p(\theta_0)=p_*,\quad q(\theta_0)=q_*,
\end{equation}
lies between the integral curves of the equations
\begin{equation}\label{q_pK}
\dfrac{d \bar p}{d q}=\frac {-q+1-e_0}{K_\pm \bar p}, \quad \bar p(\theta_0)=p_*,\quad q(\theta_0)=q_*,
\end{equation}
which have the same initial conditions.

The equations \eqref{q_pK} are easily integrated
\begin{equation}\label{K}
K_\pm \bar p^2+q^2 -2 (1-e_0)q = K_\pm p_*^2+q_*^2 -2 (1-e_0)q_*,
\end{equation}
which generates the equations of the ellipses.

At time zero, the motion starts clockwise from the point $(q=1, p=p_0)$, and the solution loses smoothness at the moment $q=0$. The trajectory $L$, which bounds the integral curve \eqref{q_p} and controls the vanishing of $q$, is glued together from pieces of the integral curves \eqref{K}, denoted by $L_\pm$. The gluing points are located where the expression $\frac {-q+1-e_0}{ p}$ changes sign, that is, where $\bar p=0$ or $\dfrac{d \bar p}{d q}=0$. The vanishing of $q$ can only occur in the lower half-plane $\bar p<0$.


For simplicity, we  assume that $p_0=0$. Then, in the lower half-plane, the lower limiter is initially the curve $L_-$, which at the minimum point, where $q=1-e_0$, is glued to the curve $L_+$. The starting point for $L_+$ should be the point $q_1=1-e_0$, $p_1=-\sqrt{\frac{e_0^2+K_-p_0^2}{K_-}}$ (it is found from \eqref{K}). Next, again from \eqref{K} (with the starting point replaced by $(q_1, p_1)$), we must find where $L_+$ intersects the axis $\bar p=0$. If the smaller root of the resulting quadratic equation for $q$
$\left(q= 1-e_0-\sqrt{\frac{e_0^2+K_- p_0^2}{K_-}}\right)$ is positive, then the integral curve lies in the upper half-plane, meaning that the lifetime of the solution is extended by one more turn and is estimated from below as $2 \pi$.
A sufficient condition for this is
\begin{equation}\label{cond1}
p_0^2 + 2\, e_0-1< \left(1-\frac{1}{ K_-}\right) e_0^2 \quad (<0).
\end{equation}
The right-hand side of \eqref{cond1} is negative, so this condition is more stringent than the sufficient condition for preserving smoothness in the non-relativistic case~\cite{RCh2021}. The smaller the amplitude of $P$, the closer $K_-$ is to 1.

 If the phase trajectory starts from the upper half-plane, then it spends some time, less than half a period, in the upper half-plane, and upon entering the lower half-plane, it may no longer satisfy the condition \eqref{cond1} and, consequently, may intersect the line $q=0$ during a time less than half a period. Then the total lifetime of the solution will be greater than $\pi$, but less than $2\pi$. In this case, we must perform the procedure of gluing the estimating curve, starting from the upper half-plane, and obtain a condition that this curve will again end up in the upper half-plane, that is, the solution will exist for a time of $2\pi$. This condition has the form
  \begin{equation*}\label{cond2}
p_0^2 <K_-(1-e_0)^2-\frac{ e_0^2}{K_-}.
     \end{equation*}

\begin{center}
\begin{figure}[htb]
\centerline{
\includegraphics[scale=0.22]{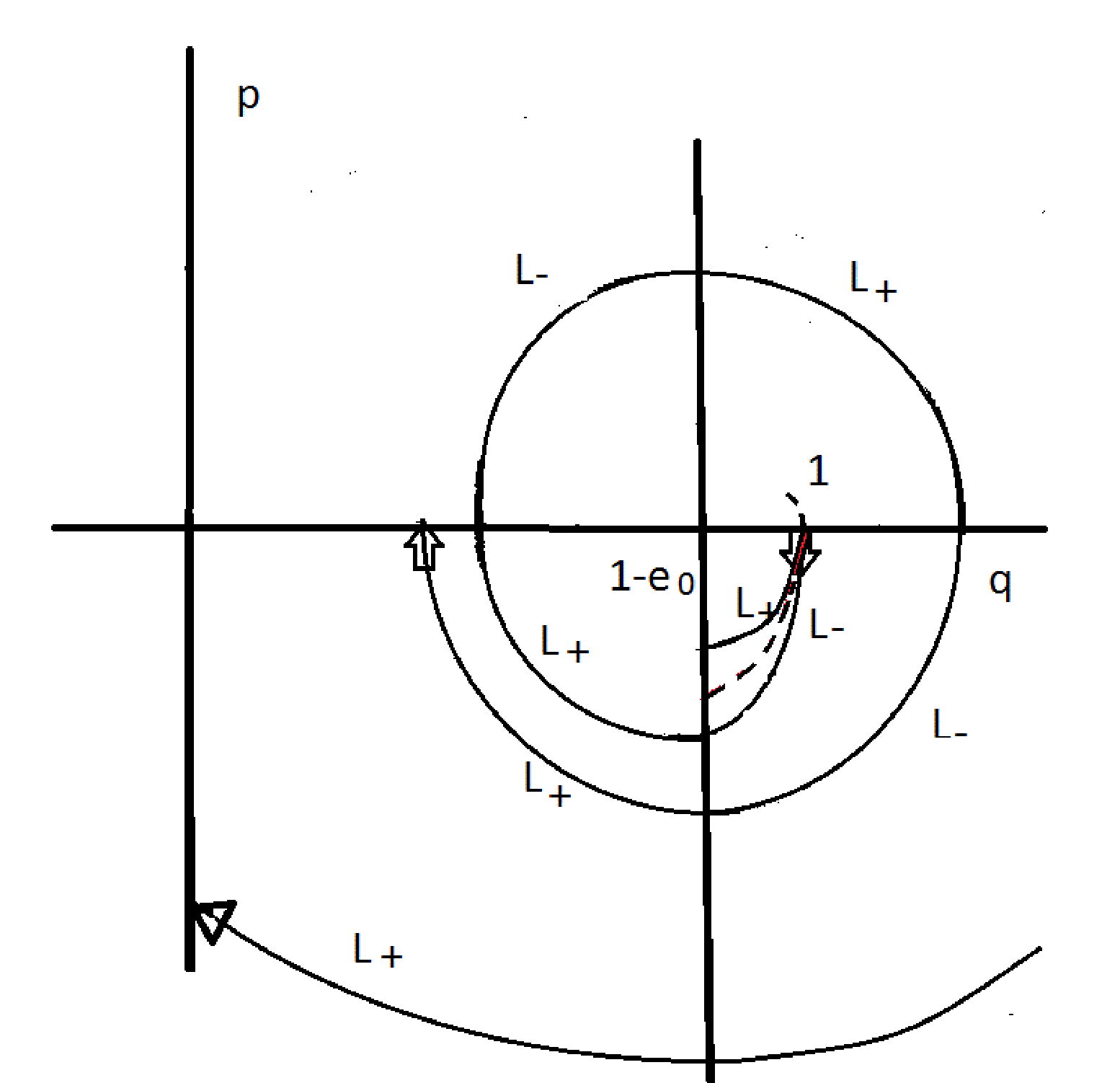}}
\caption{Limitation of the integral curve on the $(q,p)$ plane. In the first step of the evaluation, the upper ($L_-$) and lower ($L_+$) limiters are shown, and the integral curve itself is highlighted by a dash line. }\label{Pic1}
\end{figure}
\end{center}

    If we want to know whether $q $ will vanish on the next, second, turn, we need to perform four more gluings, alternating $L_+$ and $L_-$ at each step and choosing the starting points using the algorithm described above, each time relying on the expression \eqref{K} and taking the gluing point as the starting point. The gluing procedure can then be repeated any number of times.
The position of the extrema of the ellipses will always be at the point with the ordinate $q=1-e_0$, and the absolute value of $\bar p$ at the extremum point is multiplied by $\frac{1}{K_-}>1$ with each step (see Fig. \ref{Pic1}).

 Such a construction results in the proof of Theorem \ref{T1}.

\subsection{Comments to Theorem \ref{T1} }

Let us make a few comments on the result obtained:

1) Note that the right-hand side of \eqref{condn} remains positive for $K_-$ sufficiently close to unity for all finite values of $n$;

\medskip

2) It is easy to verify that for $K_-=1$, i.e., for the case of non-relativistic oscillations, Theorem \ref{T1} gives
the known condition (criterion) for the preservation of smoothness \cite{RCh2021},
\begin{equation*}
p_0^2 + 2\, e_0-1<0.
\end{equation*}

\medskip

3) The condition \eqref{condn} is only sufficient and may be far from necessary. However, calculations show that replacing it with a condition guaranteeing the smoothness of the solution for the non-relativistic case does not ensure smoothness in the relativistic case, even on the interval $[0, \pi]$.

\medskip

4) From the method of proving Theorem \ref{T1}, it follows that when estimating the lifetime of a smooth solution from below, the accuracy is lost with each subsequent revolution of the phase curve. Therefore, we obtain the most accurate estimate of the overturning time for $n=1$. This is confirmed by numerical calculations, see below.

\medskip

5) To understand how far the lower bound for the lifetime of a smooth solution contained in
Theorem \ref{T1} is from the actual blowup time, one can only use a numerical experiment, which will be conducted in Section \ref{S4}.

\subsection{Asymptotic estimate of the blow-up time for small initial disturbances}

It should be noted that the condition \eqref{condn} can be useful for asymptotic estimation of the blow-up time
of relativistic plasma oscillations. Recall that according to~\cite{RCh2021}, if
the expression $2\sqrt{1+P_0^2}+E_0^2$ is not identically equal to a constant, then, for any arbitrarily small initial data, the derivatives of the solution to the Cauchy problem \eqref{u1}, \eqref{cd1},
tend to infinity in finite time.
This is a special case of a more general result, which applies to arbitrary, and not just small, initial data \cite{RozPhD2024}.

Let us consider the case of initial perturbations in which only the electric field  differs from zero. This property is typical, in particular, for initial data traditionally used for numerical experiments simulating the effect of an electric pulse with a Gaussian potential on plasma~\cite{FrChPhScr20}:
\begin{equation}\label{cd1p}
     P_0(\rho) = 0, \quad
     E_0(\rho) = \left(\dfrac{a_*}{\zr_*}\right)^2 \zr\, \exp\{-2 \zr^2/\zr^2_*\}, \quad
 \rho \in {\mathbb R},
 \end{equation}
$ {a_*}={\rm const}>0, {\zr_*}={\rm const}>0. $

Let the condition \eqref{condn} be satisfied for some $n$. Since there are points $(\theta_0, \zr_0)$ for which $\Phi(\theta_0, \zr_0)=
\left[K_-^{n-1}(1-e(\theta_0, \zr_0))^2-\frac{ e(\theta_0, \zr_0)^2}{K_-} \right]\le 0$, so there is an infimum over all $\theta_0>0$ included in the pair $(\theta_0, \zr_0)$ for which $\Phi(\theta_0, \zr_0)> 0 $. We denote the value of this infimum by $\theta_*$ and take it as the new initial moment of time. Thus, for $\theta=\bar \theta$, there exists a point $\bar \zr \in \mathbb R$ at which
\begin{equation*}
K_{-}^n(1-e_0)^2 - e_0^2 = 0,
\end{equation*}
where $K_{-} = 8/\left(2+E_0^2\right)^3, \; e_0 = E_0'.$
Thus, we will construct initial data that are as close as possible to those at which the condition \eqref{condn} is satisfied.
If we assume that
$E_0$ can be considered a small parameter, then $n$ can be approximately expressed in terms of $E_0$ and $e_0$.
Namely,
\begin{equation*}
1 - e_0 = e_0 \left( 1 + \dfrac32 \, E_0^2 + \dots \right)^{n/2},
\end{equation*}
  therefore
 \begin{equation*}
 n \approx \dfrac43 \, \dfrac{1-2 e_0}{E_0^2 \, e_0}.
\end{equation*}

Note that this implies, in particular, that as both the amplitude of the small perturbation and its derivative decrease, the value of $n$ increases. Moreover, for $n$ to increase, it is important that the entire expression $E_0^2 e_0$ tend to zero, rather than its individual factors.

By Theorem \ref{T1}, for small deviations from the equilibrium position, we can estimate the time during which the solution remains smooth as
$$
t_{*} > n\, \pi \approx \pi \,\dfrac43 \, \dfrac{1-2 e_0}{E_0^2\, e_0},
$$
which in turn allows us to estimate the order of the breakup time of small oscillations as
\begin{equation*}\label{prop}
\zt_{br} \sim
\dfrac{1}{E_0^2(\bar \zr)} \, \dfrac{1}{1 - N_0(\bar \zr)},
\end{equation*}
which coincides with the order of the blow-up time of oscillations obtained in~\cite{FrChPhScr20}
by asymptotic methods for weakly nonlinear oscillations under the initial conditions \eqref{cd1p}.

\section{Numerical experiment} \label{S4}

Note that fixed initial data \eqref{cd1} for equations \eqref{u1} can satisfy condition \eqref{condn}
simulta\-neously for several values of $n$. Therefore, natural questions arise about estimating the smoothness time of the solution generated by
different $n$, and simul\-taneously about finding the best $n$ that can yield the most accurate prediction of the
actual  blow-up time.

To answer these questions, we conduct a computational experiment. We fix the initial data \eqref{cd1} in the form
\begin{equation}\label{cd1all}
E_0(\zr) = \za \, \zr\, \exp\{-2 \zr^2/\zr^2_*\}, \quad P_0(\zr) = \zb \, \zr\, \exp\{-2 \zr^2/\zr^2_*\},
\end{equation}
where the quantities $\za, \, \zb$ and $\zr_*$ characterize the amplitude and width of the laser pulse that initializes the oscillations.
Note that similar perturbations in the plasma can be obtained using a cylindrical lens in full-scale experiments~\cite{Shep13}.

Then, using the high-precision algorithm~\cite{RChGVM21}, we numerically solve the extended system generated by the statement
\eqref{u1}, \eqref{cd1}. This yields the coordinates of the blow-up point $(T_{br}, \, \zr_{br})$
with a predetermined accuracy (in the calculations, an accuracy of about $10^{-10}$ was used, and the blow-up occurred at two points located
symmetrically relative to the origin).

During the numerical experiment we find
$$P(\zt,\zr), \,p(\zt,\zr), \,E(\zt,\zr), \,e(\zt,\zr)$$  for  $n = 1, 2, 3$, the left-hand side of  condition \eqref{condn}.
It was assumed that the current (for an arbitrary point in time) smooth solutions  could be used as initial conditions.

As a result, for each calculation variant, the point in time $\zt_n$ was determined, depending on $n$, at which the sign of the condition \eqref{condn} changed from positive to negative. This procedure allows us to estimate the predictive capabilities of the parameter $n$. Finally, using the formula $T_{n,sm} = \zt_n + n\, \pi$, the predicted time corresponding to $n$ was calculated. Then we compare this time with the actual moment of blow-up $T_{br}$, which was determined solely by the initial conditions.
Next, from the predicted smoothness preservation time $T_{n,sm}$, we select the time closest to $T_{br}$.

The calculations used discrete time and space steps of the order of $10^{-3}$, and time integration was performed using the classical Runge-Kutta method of fourth-order accuracy.

We present the results of the numerical experiments.

$$
\begin{array}{c}
1) \; \za = 0.4761, \; \zb = 0, \; \zr_* = 3, \; T_{br} = 55.106 \dots,\; \zr_{br} = \pm 0.19376 \dots     \\
T_{1,sm} \approx 54.34  , \quad  T_{2,sm} \approx 51.68  , \quad  T_{3,sm} \approx 51.32;
\end{array}
$$

\begin{center}
\begin{figure}[htb]
\centerline{
\includegraphics[scale=0.8]{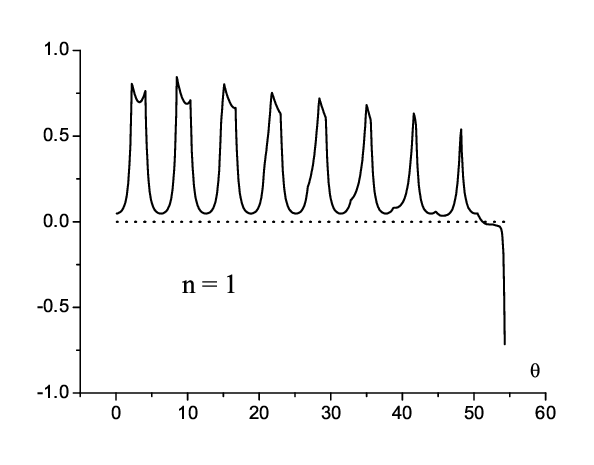}}
\caption{Dynamics of the sufficient condition \eqref{condn} for $n=1$: the solid line represents the value on the left side of the condition as a function of the initial time, and the dotted line represents the background zero value.}\label{Pic2}
\end{figure}
\end{center}

For calculation variant $1)$, Fig. \ref{Pic2} shows the dynamics of the sufficient condition   for $n=1$: the solid line denotes  values of the expression on the left-hand side of \eqref{condn} as a function of the initial time, and the dotted line denotes the zero background value.
The intersection of the graphs at time $\zt_1 \approx 51.2$ means that, in accordance with the sufficient condition \eqref{condn}
if we take the solution to problem \eqref{u1} at an arbitrary time $\zt \le \zt_1$ as the initial condition \eqref{cd1}, then
the solution will remain smooth and bounded, along with its derivatives, over a time interval of length no less than $\pi$.

\begin{center}
\begin{figure}[htb]
\includegraphics[scale=0.8]{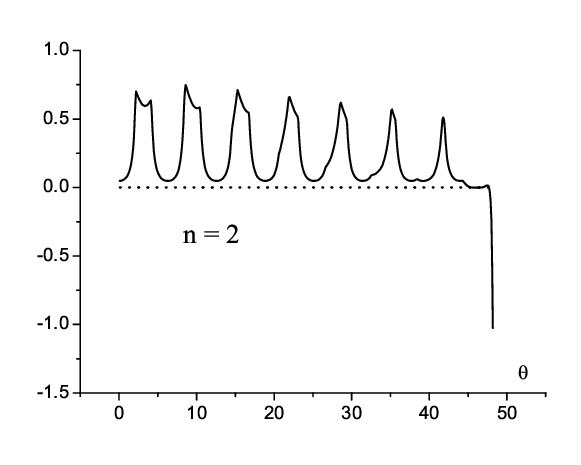}
\caption{Dynamics of the sufficient condition \eqref{condn} for $n=2$: the solid line denotes the values of the expression on the left-hand side of \eqref{condn} as a function of time, and the dotted line denotes the background zero value.}\label{Pic3}
\end{figure}
\end{center}

Similar graphs for $n=2$ are shown in Fig. \ref{Pic3} for comparison purposes for the same calculation variant $1)$.
In turn, the intersection of the graphs at time $\zt_2\approx 45.4$ means that, in accordance with the sufficient condition \eqref{condn}
if we take the solution of problem \eqref{u1} at an arbitrary time $\zt \le \zt_2$ as the initial condition \eqref{cd1}, then
the solution will remain smooth and bounded, along with its derivatives, over a time interval of length no less than $2 \pi$.

Figures \ref{Pic2} and \ref{Pic3} illustrate the typical behavior of the condition \eqref{condn}: while the solution to the problem remains smooth, the left-hand side of the condition is positive. Then, a transition to the region of negative values ​​signals an impending loss of smoothness. The length of the segment of remaining smoothness of the solution
is a multiple of $\pi$. The following calculation examples involve varying parameters in the initial conditions \eqref{cd1all}.

$$
\begin{array}{c}
2)\;  \za = 0.4761, \; \zb = 0, \; \zr_* = 4.5, \; T_{br} = 29.493 \dots,\; \zr_{br} = \pm 0.13653 \dots     \\
T_{1,sm} \approx 25.84  , \quad  T_{2,sm} \approx 23.08  , \quad  T_{3,sm} \approx 22.52;
\end{array}
$$

 $$
\begin{array}{c}
3)\;  \za = 0.4761, \; \zb = 0, \; \zr_* = 6, \; T_{br} = 16.999 \dots,\; \zr_{br} = \pm 0.20376 \dots     \\
T_{1,sm} \approx 16.14  , \quad  T_{2,sm} \approx 12.98  , \quad  T_{3,sm} \approx 12.02;
\end{array}
$$
In the calculation variant $3)$ for $n=3$ at the initial moment of time the expression on the left side of \eqref{condn} was negative, then became positive, and finally, at $\zt_3 \approx 2.6$ it finally changed from positive to negative values.

It is easy to see that for $\zb=0$ the most accurate predictive property for the value of the solution smoothness interval is the condition \eqref{condn} for $n=1$. We now present the calculation results for variants of the initial conditions for $\za =0$.

$$
\begin{array}{c}
4) \; \za = 0, \; \zb = - 0.6129 \; \zr_* = 4, \; T_{br} = 34.852 \dots,\; \zr_{br} = \pm 0.45274 \dots     \\
T_{1,sm} \approx 28.04  , \quad  T_{2,sm} \approx 24.78  , \quad  T_{3,sm} \approx 24.02;
\end{array}
$$

$$
\begin{array}{c}
5) \; \za = 0, \; \zb = - 0.7857 \; \zr_* = 4, \; T_{br} = 15.929 \dots,\; \zr_{br} = \pm 0.47745 \dots     \\
T_{1,sm} \approx 14.94  , \quad  T_{2,sm} \approx 14.38  , \quad  T_{3,sm} \approx 10.42;
\end{array}
$$

$$
\begin{array}{c}
6) \; \za = 0, \; \zb = - 0.9088 \; \zr_* = 4, \; T_{br} = 15.023 \dots,\; \zr_{br} = \pm 0.10343 \dots     \\
T_{1,sm} \approx 11.24  , \quad  T_{2,sm} \approx 7.38  .
\end{array}
$$

In the calculation variant $6)$, for $n=3$ and $\zt \ge 0$, the expression on the left-hand side \eqref{condn} was negative, which
prevented us from predicting the smoothness interval of the solution $T_{3,sm}$.

It is easy to see that for $\za=0$, the most accurate predictor of the smoothness interval of the solution is the condition \eqref{condn} for $n=1$.

To fully describe the process of smooth solution failure, we provide
additional illustrations. For this purpose, we will again use the parameters of variant $1)$.

\begin{center}
\begin{figure}[htb]
\includegraphics[scale=0.8]{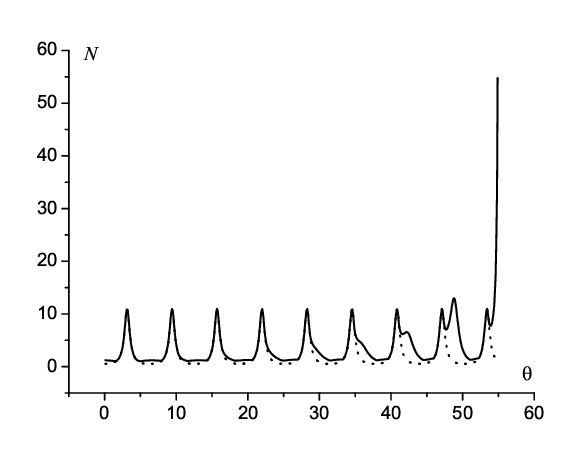}
\caption{Electron density dynamics: maximum over the region (solid line)
and at the origin (dashed line).}\label{Pic4}
 \end{figure}
\end{center}

Since the initial conditions, by virtue of  Theorem \ref{T1}, ensure the existence of a solution for more than one period, then, according to theoretical analysis, two trends can be observed in the oscillation process. The first is that density oscillations outside the origin are slightly ahead in phase of density oscillations at $\rho = 0$, and this phase shift increases from period to period. The second trend is more obvious: over time, an absolute density maximum, located outside the origin, gradually develops.
In Fig. \ref{Pic4}, the dotted line shows
the change in electron density at the origin over time, and the solid line shows the dynamics of the maximum value over the domain.
Initially, the oscillations are regular, i.e., global density maxima and minima over the domain alternate
every other after half a period and are located at the origin. After the seventh regular (central)
maximum at time $\theta \approx 42.2$, a new structure emerges, namely, an electron density maximum outside the origin, while regular oscillations continue to be observed in the vicinity of the origin. The newly emerged maximum, at time $\theta \approx 48.8$, approximately doubles in magnitude, and in the next period at $\theta \approx 55.1$ a delta-shaped electron density singularity emerges in its place.

\subsection{Qualitative properties of the solution near the blow-up point}

\begin{center}
\begin{figure}
\includegraphics[scale=0.8]{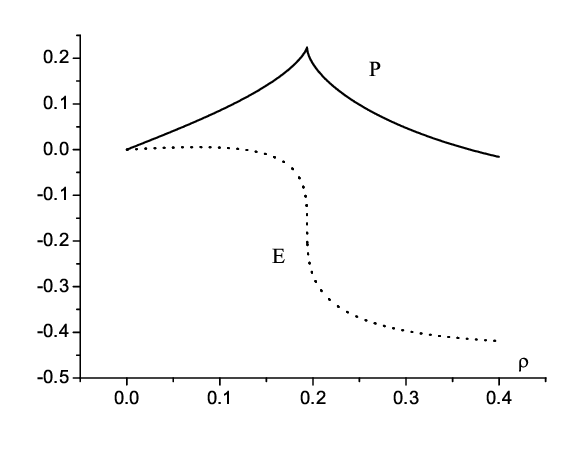}
\caption{
Spatial distribution of momentum and electric field at the moment of blow-up.}\label{Pic5}
\end{figure}
\end{center}

Fig. \ref{Pic5} shows the spatial distributions of the momentum $P$ and the electric field $E$ at the instant $\theta \approx 55.1$, when
the "gradient catastrophe" occurred, i.e., the functions $P$ and $E$ themselves remained bounded, but their spatial derivatives did not.
Note that, due to the structure of the equations \eqref{u1}
its solution will remain an odd function of the coordinates if the initial data have this property. The initial data
\eqref{cd1all} are precisely such, so we present the results of numerical calculations only on the positive semi-axis.

\begin{center}
\begin{figure}
\includegraphics[scale=0.8]{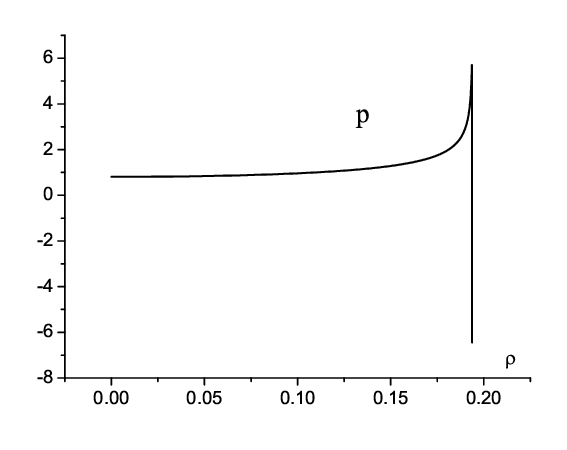}
\caption{
Spatial distribution of the derivative of the momentum $p$ in the vicinity of the blow-up point (macro).}\label{Pic6}
\end{figure}
 \end{center}

Let us note the fundamentally different behavior of the momentum and the electric field in the vicinity of the density singularity.
The electric field function develops a strong discontinuity (a "step"), and its derivative is monotonic, eventually
approaching minus infinity. It may seem that the momentum function develops a weak discontinuity, i.e., its derivative jumps from
from positive to negative values. This is indicated, for example, by Fig. \ref{Pic6}, which depicts the momentum derivative
in the vicinity of the blow-up point.

\begin{center}
\begin{figure}
\includegraphics[scale=0.8]{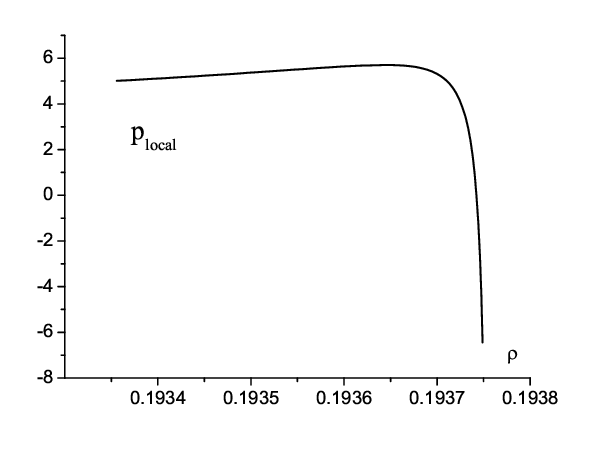}
\caption{
Spatial distribution of the derivative of the momentum $p$ in the vicinity of the blow-up point (micro).}\label{Pic7}
\end{figure}
 \end{center}

However, a more detailed examination reveals something else. As shown in Fig.\ref{Pic7}, the momentum derivative does indeed transition
from positive to negative values, but not abruptly, but smoothly, with a local maximum. And after that,
as it approaches the singularity point, it monotonically approaches negative infinity, just like the electric field function.

This observation can be justified as follows.
Consider the second equation \eqref{char1d}:
\begin{equation*}
- \dfrac{d \ln |1-e|}{d\theta}=\frac{p}{(1+P^2)^{3/2}}.
\end{equation*}
Let $\zt_*$ be the formation time of the singularity, $\zt<\zt_*$. Integrating, we obtain
\begin{equation*}
- \ln |1-e(\zt_*)| + \ln |1-e(\zt)|=\int\limits_{\zt}^{\zt_*}\frac{p}{(1+P^2)^{3/2}} d t \sim C \int\limits_{\zt}^{\zt_*}\,{p}\, dt,\quad C>0.
\end{equation*}
If $e(\zt_*)\to-\infty$, then $p(\zt_*)\to-\infty$.

We emphasize that in the vicinity of the blow-up point, the bounded functions of momentum and electric field behave differently:
the electric field $E$ tends to a stepwise (discontinuous) function with a monotonic derivative tending to negative infinity, while the momentum $P$
forms a continuous "corner" profile with a continuous (smooth) transition of the derivative from positive to negative values.
As the spatial singularity of density $N$ is approached, the negative derivative of the momentum also tends to negative infinity.

\section{Conclusion}\label{S5}
Let us discuss the applicability of the obtained
results. Examples are given in the work~\cite{FrChPhScr20} of how the parameters of a laser pulse propagating in a plasma influence the breaking of plasma oscillations. In particular, the breaking effect, when choosing dimensionless quantities as in the calculation variant $1)$, can be easily realized under laboratory conditions, i.e., it corresponds to the actual parameters of a laser pulse.
In turn, the sufficient condition obtained using analytical methods in this work adequately describes the duration of smooth oscillations, which can be reliably established using numerical experiments. Taken together, this means that the presented theoretical result
may be useful in specific (determined) practical studies.

It is also worth noting the "serial" (or "parametric") nature of condition  \eqref{condn}, namely: for a fixed set of initial conditions,
a whole series of smoothness intervals, multiples of $\pi$, are simultaneously established. In other words, the researcher has a choice of the duration
of the interval of sufficient solution smoothness, based on certain additional considerations. For example, solution smoothness is often required to justify the convergence of approximate solutions to exact ones.

\section*{Acknowledgment}
O.R. was supported by Russian Science Foundation  grant 23-11-00056.

\section*{Declarations}
The authors declare no competing interests.

\end{document}